# Operation of a Bloch oscillator

K. F. Renk[*], A. Meier, B. I. Stahl, A. Glukhovskoy, M. Jain, H. Appel, and W. Wegscheider

*Institut für Angewandte Physik, Universität Regensburg, 93040 Regensburg, Germany*

**We report the operation of a Bloch oscillator. The active medium was a static-voltage driven, doped GaAs/AlAs superlattice which was electromagnetically coupled to a resonator. The oscillator produced tuneable microwave radiation (frequency ~ 60 GHz; power ~ 0.5 mW; efficiency ~ 4 %). The gain (~ $10^4$ cm$^{-1}$) was due to the nonlinearity mediated by miniband electrons. We also present a theory of the oscillator. The Bloch oscillator should in principle be feasible for generation of radiation up to frequencies of 10 THz and more.**

[*] *karl.renk@physik.uni-regensburg.de*

Conduction electrons in a semiconductor superlattice can undergo Bragg reflections at the superlattice planes and the energy of motion along the superlattice axis can be confined to a miniband [1, 2]. The confinement can cause a negative differential resistance [2] and Bloch oscillations [3] which have been observed by transport [4] and optical [5] studies, respectively. Ktitorov et al. [6] presented a theory indicating that a superlattice in a negative-differential resistance state should be a ("Bloch") gain medium for high frequency radiation from almost zero up to the Bloch frequency, which is determined by the superlattice period and the strength of a static field applied to the superlattice and can reach 10 THz or more. Gain has been concluded [7] from an anomalous THz transmissivity of an array of superlattices, switched by a voltage pulse into a negative-differential resistance state. In this Letter, we report the operation of a Bloch oscillator, i.e. an oscillator based on Bloch gain, with a voltage-driven semiconductor superlattice as the active medium coupled to a resonator. We also present a theoretical description of the oscillator properties.

In the Bloch oscillator (Fig. 1a), a superlattice which is part of a superlattice electronic device (SLED) is connected to a static-voltage source delivering a current $I$. The superlattice generates, by stimulated emission, radiation at the resonance frequency, $\nu$, of the resonator.

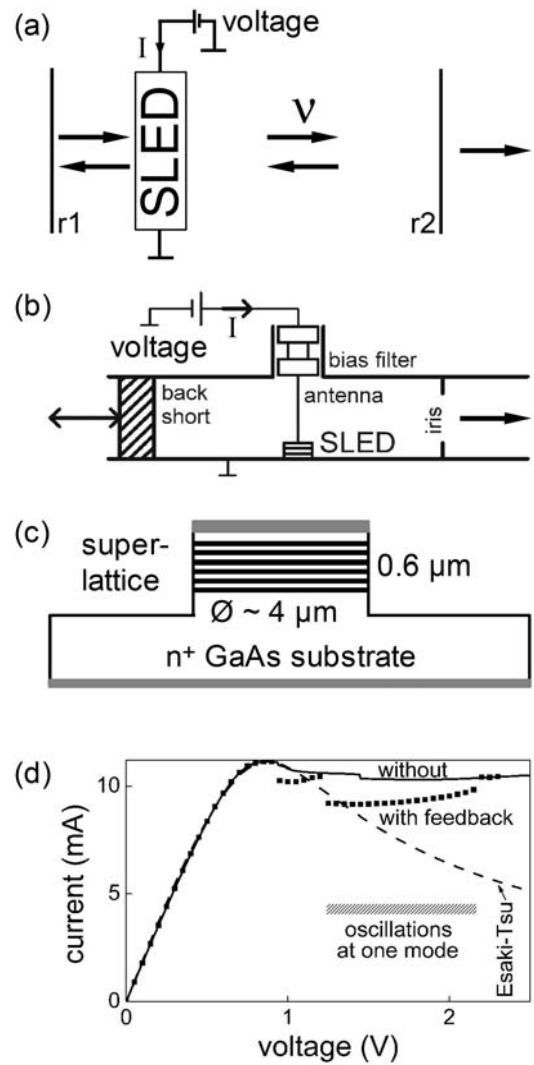

Fig. 1. (a) Principle of the Bloch oscillator; r1, reflector and r2, partial reflector. (b) Arrangement. (c) Superlattice electronic device (SLED). (d) Current-voltage curve of the SLED.

By varying the resonator length, $\nu$ is changed. In our arrangement (Fig. 1b), a SLED was mounted in a metal-cavity resonator (height 2 mm, width ~ 4mm) with a moveable backshort. A gold whisker antenna coupled the SLED to the resonator and connected it to a voltage source. A filter avoided radiation loss to the bias circuit. By changing the diameter of the iris, we varied the quality factor of the resonator.

The SLED (Fig. 1c) contained the superlattice (130 periods, each period consisting of 14 monolayers of GaAs and 2 monolayers of AlAs) embedded in gradual layers and $n^+$ GaAs layers, grown by molecular beam epitaxy on an $n^+$ GaAs substrate. Top and bottom of the SLED were covered with ohmic contact layers. The current-voltage curve of the SLED (Fig. 1d, solid line) was ohmic at small voltage. The current showed, with increasing voltage, a maximum at the peak current $I_p$ (~ 12 mA; peak-current density $j_p$ ~ 100 kA/cm$^2$) and then decreased slightly.

Figure 2 exhibits an emission line (at a frequency near 60 GHz) of the oscillator. The line broadening was mainly due to an insufficient stabilization of our voltage source and to thermal fluctuations within the SLED; with a 1.5V battery, the halfwidth was about 200 kHz. Outside the line center, the signal decreased strongly; the background was mainly due to the noise of the spectrum analyzer we used to monitor the line. A thermal power meter indicated a power (0.5 mW) which corresponded to an efficiency of about 4 percent for conversion of electric to radiation power.

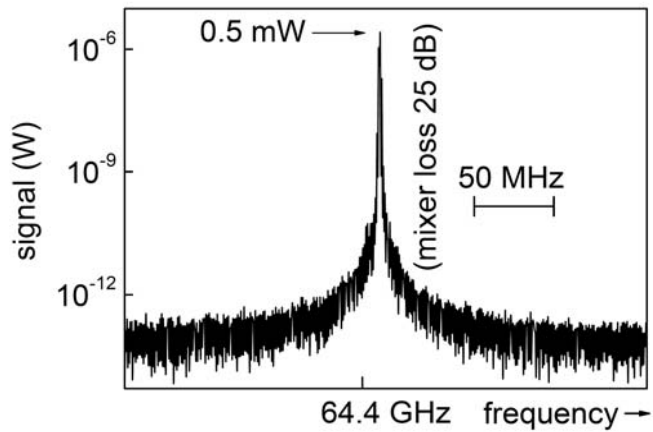

Fig. 2. Emission line of the Bloch oscillator.

The power was obtained for a reflectivity of the iris of about 0.7. From this we conclude that the gain coefficient of the active superlattice was about $10^4$ cm$^{-1}$. To study the tuning behavior, we used an iris of smaller diameter. We found that mechanical tuning was possible over a wide frequency range (Fig. 3a), however, $P/P_{max}$ (Fig. 3a, upper part), which is the power relative to that of maximum emission, decreased strongly if the resonator length deviated from that of maximum power $P_{max}$. The current increased slightly (by few percent). We observed that the oscillation frequency $\nu$ increased with increasing voltage (Fig. 3b). Within a voltage range (1.2 V to 2.2 V) in which the oscillator operated at one mode, $\nu$ increased by about 1 percent; the power increased by a factor of two. An increase of current (Fig. 1d, dotted line) indicated a positive differential resistance of the superlattice in the oscillating state. The switching to an oscillating state was joint with an abrupt decrease of current from a point on the solid curve to the corresponding point on the dotted curve.

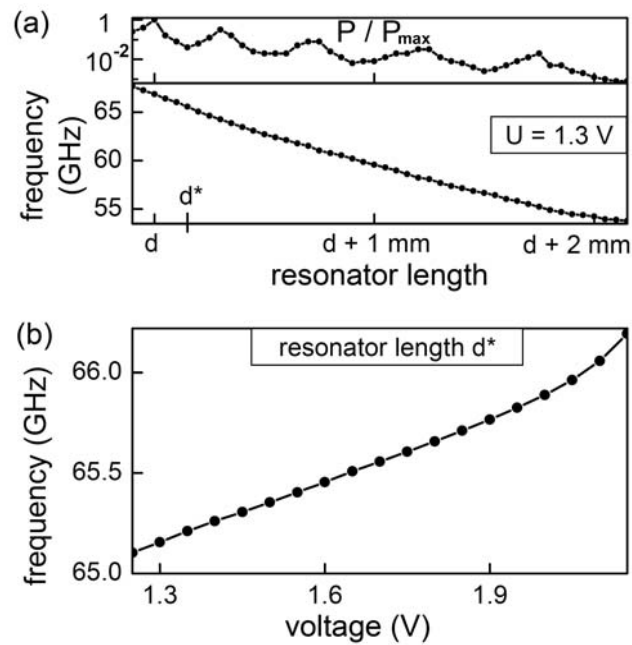

Fig. 3. (a) Frequency (and power) of the Bloch oscillator for different resonator lengths; d, resonator length of maximum power. (b) Frequency of the Bloch oscillator for different values of the voltage.

We attribute the oscillation to Bloch gain. For an analysis of our results, we use the dispersion relation $\varepsilon = \frac{1}{2}\Delta(1-\cos ka)$ where $\varepsilon$ is the energy, $k$ the wave vector of an electron moving along the superlattice axis, $\Delta$ the miniband width and $a$ the superlattice period. We describe the electron as a wave packet with its center given by the trajectory $\xi$. Under the action of a static field, $E$, the electron performs a periodic motion (Fig. 4a) between $-\hat{\xi}$ ($\varepsilon = 0$) and $+\hat{\xi}$ ($\varepsilon = \Delta$). The corresponding de Broglie wavelength changes from $-\infty$ to $2a$, respectively. When the de Broglie wavelength reaches the value $2a$, the electron is Bragg reflected and reverses its direction. Accordingly, the electron oscillates with the Bloch frequency $v_B = \frac{1}{h}eaE$ where e is the elementary charge and h Planck's constant. Under the additional action of a high frequency field at the frequency $v = v_B$, the electron transfers, within one half period, energy to the high frequency field. During the other half period, the same energy is transferred back to the electron. If relaxation is introduced, there is a net energy transfer via the Bloch oscillating electrons from the static to the high frequency field if $v < v_B$ and vice versa if $v > v_B$. The corresponding solution of the Boltzmann equation delivers the time dependent drift velocity [8]

$$v(t) = 2v_p \int_{-\infty}^{t} \frac{dt_0}{\tau} \exp\left(-\frac{t-t_0}{\tau}\right) \cdot \sin\left[-\int_{t_0}^{t} \frac{ea}{\hbar} E(t_1) dt_1\right] \qquad (1)$$

where $v_p = \frac{1}{4\hbar}\Delta a$ is the peak-drift velocity, $\tau$ the intraminiband relaxation time, $E(t) = \frac{1}{L}(U + \hat{U}\cos \omega t)$ the instantaneous field strength, and $L$ the superlattice length; for simplicity, we neglect elastic scattering at defects.

Without feedback ($\hat{U} = 0$), eq.(1) delivers the Esaki-Tsu characteristic [2] $I = 2I_p(U/U_c)(1 + U^2/U_c^2)^{-1}$ where is the critical voltage, i.e. the voltage across the superlattice at the peak current $I_p = NeAv_p$ where N is the free carrier concentration and A the superlattice cross section area. Taking into account a series resistance (Rs), our experimental curve is reproduced by the theory (Fig. 1d, dashed) in the range of positive differential resistance if we choose appropriate data (Rs ~ 30 Ω; Uc ~ 0.57 V; N ~ 2.3·10$^{16}$ cm$^{-3}$), and τ (~ 1.5·10$^{-13}$ s). The corresponding critical field $E_c$ = $U_c/L$ ~ 10 kV/cm and $v_p$ (~ 10$^7$ cm/s) are in accordance with the microscopic parameters (Δ ~ 0.14 eV; L ~ 0.6 μm; a ~ 4.3 nm).

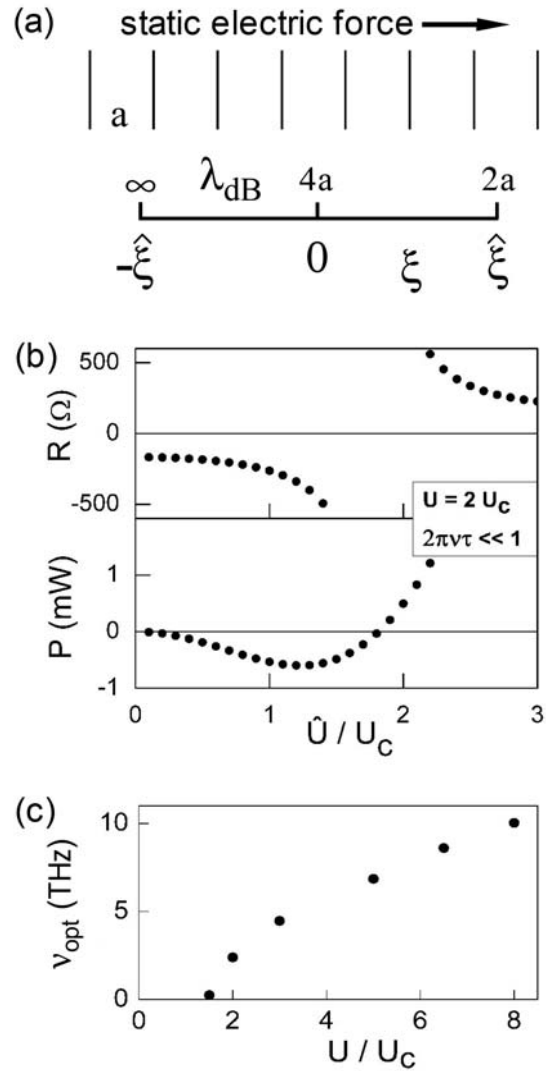

Fig. 4. (a) Motion of an electron in a superlattice along the superlattice axis; ξ, trajectory and λ$_{dB}$, de Broglie wavelength. (b) Resistance of the superlattice (upper part) and power (lower part) for different amplitudes Û of the high frequency voltage. (c) Optimum-gain frequency, ν$_{opt}$, for different values of the static field.

We determined, for a fixed static voltage ($U = 2\, U_c$) and different values of $\hat{U}$, the amplitude $\hat{v}$ of the drift velocity at the frequency $v$ and of the current ($\hat{I}/I_p = \hat{v}/v_p$). Then, we calculated the dynamic resistance $R = \hat{U}/\hat{I}$ and the corresponding power $P = \frac{1}{2}\hat{U}^2/R$. There is a range of negative values of $R$ and $P$ (Fig. 4b) indicating gain (for $v < v_B$) and another region of positive values indicating absorption ($v > v_B$). $|R|$ increases with increasing amplitude of the high frequency field.

This shows that the superlattice can match itself to the external circuit. There is an optimum operation point for $\hat{U}$ (at $R \sim -300\,\Omega$) where a maximum power can be transferred to the high frequency field. The maximum power ($0.1 \cdot I_p U_c \sim 0.6$ mW) is almost equal to the experimental value. The frequency, $v_{opt}$, of optimum gain increases with the static field strength (Fig. 4c) and can reach (for $U \sim 8\, U_c$) a value of 10 THz. The observation of a frequency increase (Fig. 3b), though limited by the resonator, is in accordance to the theoretical result. We note that our results are in accordance with earlier gain calculations [6, 9].

Theoretical studies [10] predicted the possibility that propagating space charge domains are formed in a superlattice in a negative resistance state. For a corresponding domain-mediated oscillator, the oscillation frequency $v_{dom}$ would be equal to the reciprocal transit time, a domain takes to form and to traverse a superlattice. According to the Esaki-Tsu curve (Fig. 1d) it is expected that $v_{dom}$ decreases with increasing voltage; this is confirmed by extended simulations [11] and is also known for Gunn oscillators, which are domain oscillators. For our superlattice, the criterion of domain formation, $(NL)_d \geq 5 \cdot 10^{11}$ cm$^{-2}$ [7, 12] was fulfilled, $NL \sim 3\, (NL)_d$. The observation of an increase of the oscillation frequency with the voltage (Fig. 3b) suggests however that the Bloch-gain state governed the oscillation. Further evidence follows from the good agreement between the calculated and measured power. The absence of a jump like, strong current decrease near maximum current in the experimental current-voltage curve for the superlattice without feedback (Fig. 1d, solid line) is a direct sign of a minor role of domains.

The Bloch oscillator can be described as a parametric oscillator. The main process is the parametric interaction of a high frequency field with an elementary, single-electron Bloch oscillator, which is driven by the static field. Due to relaxation, the Bloch frequency has an uncertainty, $\delta\nu_B$, determined by the uncertainty relation $2\pi h \cdot \delta\nu_B \tau \geq h$. Accordingly, parametric interaction requires only the weak criterion $n\nu \sim \nu_B$ where n is an integer number. The gain for the parametric interaction is of the same order of magnitude from almost zero frequency up to frequencies well above $2\pi\nu\tau \sim 1$, e.g. $\sim 10$ (or 10 THz), as can be concluded from results of gain calculations [6, 9].

An increase of the superlattice volume together with appropriate sample cooling and resonator design should lead to higher power levels. An increase of doping by an order of magnitude may increase the gain coefficient correspondingly. The higher doping would result in an upper cut-off frequency $j_P (\eta\varepsilon_0 E_c)^{-1} \sim 10$ THz ($\eta$, dielectric constant and $\varepsilon_0$, electric field constant). Materials with smaller effective masses, like InGaAs/InAlAs superlattices, would allow reaching frequencies above 10 THz. The use of materials with a shorter relaxation time and high doping, like GaN/GaAlN superlattices, would extend the cut off frequency to almost 30 THz. The development of a Bloch oscillator for the THz range may contribute, as the quantum cascade laser [13], which is operated at a temperature near liquid nitrogen temperature, towards a development of the THz frequency range.

In conclusion, we have reported the operation of a Bloch oscillator based on Bloch gain in a semiconductor superlattice and presented a theoretical analysis of its properties. The Bloch oscillator represents a room-temperature, tuneable monochromatic radiation source suitable for generation of microwave and THz radiation.

**Acknowledgement**

The work has been supported by the Deutsche Forschungsgemeinschaft. One of us (B.I.S.) would like to thank L. Esaki for discussions (about Bloch oscillations) during the 54[th] Nobel Laureate Meeting in Lindau, June 2004.